\documentclass[12pt]{article}
\usepackage{amsfonts}
\usepackage{amsmath}
\usepackage{amscd}
\usepackage{epsfig}
\usepackage{color}
\usepackage{graphicx}

\begin{document}

{\Large\bf  Spectrum Generating Algebras for the free motion in
$S^3$.}

\vskip1cm
{M Gadella${}^\dagger$, J Negro${}^\dagger$, L M
Nieto${}^\dagger$, G P Pronko${}^\ddagger$, and M.
Santander${}^\dagger$}

\bigskip

${}^\dagger$Departamento de F\'{\i}sica Te\'orica, At\'omica y
\'Optica, Universidad de Valladolid, 47071 Valladolid, Spain \\
${}^\ddagger$Department of Theoretical Physics, IHEP. Protvino,
Moscow Region 142280, Russia.

\begin{abstract}
We construct the spectrum generating algebra (SGA) for a free
particle in the three dimensional sphere $S^3$ for both, classical
and quantum descriptions. In the classical approach, the SGA
supplies time-dependent constants of motion that allow to solve
algebraically the motion. In the quantum case, the SGA include the
ladder operators that give the eigenstates of the free Hamiltonian.
We study this quantum case from two equivalent points of view.
\end{abstract}

\section{Introduction}

The notion of spectrum generating algebra (SGA), sometimes called
non inva\-rian\-ce algebra, was introduced many years ago
\cite{BB,DGN,DO}. In the context of quantum mechanics, the idea of
the SGA consists in reducing the construction of the whole Hilbert
space for a given system to a problem of representation theory. The
knowledge of the symmetry (usually called `dynamical symmetry') of a
problem allows to solve it only partly: its representations gives
the subspace of the whole Hilbert space of eigenstates corresponding
to a fixed energy. The further extension to the SGA needs to
introduce ladder operators that change the energy, i.e., operators
that do not commute with the Hamiltonian (it is the reason to call
this construction non-invariance algebra). At the very best, the
whole set of operators ---those generating the dynamical algebra
plus the ladder operators--- may form a finite dimensional
non-compact algebra whose representation gives the Hilbert space of
the system. In this respect, the symmetry algebra of the Hamiltonian
plays the role of the Cartan subalgebra, while the additional
operators of the SGA, which do not commute with the Hamiltonian,
play the role of the Borel elements.

In the classical frame, the symmetry algebra provides constants of
motion which are functions of the dynamical variables characterizing
the possible trajectories. However, the motion is obtained from
another kind of constants of motion that include explicitly the
time. Such constants come from the elements of the SGA `not
commuting' (in the sense of Poisson brackets) with the Hamiltonian
\cite{JS}.

The main purpose of this paper is using the SGA technique to solve
the spectral problem related with the quantum Hamiltonian of the
free motion in the three dimensional sphere $S^3$, embedded in the
four dimensional coordinate space ${\mathbb R}^4$, which has a pure
discrete spectrum \cite{S}. This problem is already nontrivial,
interesting by itself and will provide important clues for the
extension of the SGA in the study of more general quantum systems
evolving on configuration spaces with constant curvature. In the
case of the free particle in $S^3$, it is well known that the
symmetry algebra is $\frak{so}(4)$ and our task is to construct the
ladder operators which do not commute with the Hamiltonian. As we
shall see below in order to achieve this goal we will need to
involve, apart from symmetry operators, also the elements of the
homogenous space of the group $SO(4)$. The main result obtained in
this work is the explicit construction of a SGA isomorphic to
$\frak{so}(4,2)$.

It is interesting to remark that this is a new method to study
classical and quantum systems in configuration spaces of nonzero
constant curvature. However, the study of such systems is not new up
to the extent that Schr\"odinger himself has obtain the levels of
energy of the Kepler problem in $S^3$, see \cite{S}. For other
approaches, see \cite{JS1,Q}.

The paper is organized as follows. Section 2 begins with the
construction of the SGA for the free particle in $S^3$ in the
classical context. This is a good starting point, as it will give us
hints for the construction of the quantum SGA of the same problem.
However, the classical version is much simpler, as is free of the
important and hampering difficulties of the ordering of
non-commuting operators, which is specific of the quantum case. In
this section we will also show how we can use the SGA in order to
solve the classical equations of motion. In section 3, we shall
present the detailed construction of the SGA for the quantum
problem. Here, we have adopted the point of view of a direct
quantization of the classical SGA and choose the representation by
means of some natural restrictive relations.  The Hilbert space of
states will be explicitly derived with the help of the SGA through
ladder operators. At the very end, we obtain position and momentum
operators along to constraint and gauge fixing relations compatible
with $S^3$. In Section 4, we adopt the opposite point of view. We
start with canonical quantizations of the classical Dirac brackets
for the variables position and momentum with their corresponding
constraint and gauge condition compatible with $S^3$.  Then
determine the SGA, as well as the ladder operators. We also derive
the restrictive relations as a consequence of our definitions.
 Finally,  we  present the concluding remarks
and some indications for future research.

\section{The classical case}

We shall start with the Lagrangian of the free motion in $S^3$,
considered as a sphere of radius one,
\begin{equation}
\label{1} L=\frac1{8{\bf x}^4}\;\sum_{ i,j=1}^4(x_i\dot x_j-\dot
x_ix_j)^2\,,\quad {\rm with}\qquad {\bf x}^2=\sum_{i=1}^4 x_i^2\,,
\end{equation}
where the dot represents the derivative with respect to time and we
have assumed $m=1$, since the mass do not play any relevant role in
our development. This Lagrangian can be considered the restriction
to the sphere $S^3$  of the free Lagrangian $L=\sum_{i=1}^4\dot
x_i^2$ defined in the ambient space ${\mathbb R}^4$. The canonical
momenta are determined by
\begin{equation}\label{2}
p_i=\frac{\partial L}{\partial\dot x_i}=\frac1{2{\bf
x}^4}\;\sum_{j=1}^4(x_j\dot x_i-x_i\dot x_j)x_j\, ,
\end{equation}
and satisfies the primary constraint

\begin{equation}\label{3}
 {\bf x}{\bf p}=x_ip_i=0\,,
\end{equation}
${\bf  x}{\bf  p}=x_ip_i=0$, where here and throughout the paper
the convention of summation over repeated indices is used (in this
respect, Latin subindexes $i,j,k,\dots$ will run from $1$ to $4$,
the dimension of the ambient space).

The Legendre transformation of the Lagrangian (\ref{1}) gives the
canonical Hamiltonian
\begin{equation}\label{4}
H=\frac12\;J_{ij}J_{ij}\,,\qquad J_{ij}=x_ip_j-x_jp_i\,,
\end{equation}
where $J_{ij}$ has the structure of an angular momentum. Our
strategy to work in $S^3$ will be the following. Instead of dealing
in the 8-dimensional phase space with dynamical variables $x_i,p_i$
satisfying the canonical Poisson brackets, we impose the gauge
fixing condition
\begin{equation}\label{2.5}
{\bf  x}^2=x_ix_i=1,
\end{equation}
and the primary constrain (\ref{3}). According to the usual
procedure \cite{D,SM}, we also introduce the Dirac brackets
\begin{equation}\label{dirac}
\{x_i,x_j\}_D=0\,,\quad    \{p_i,x_j\}_D=\delta_{ij}-x_ix_j\,, \quad
\{p_i,p_j\}_D=J_{ij}\,.
\end{equation}
In the sequel, we prefer to use the variables $x_i$ and $p_i$
subject to the Dirac brackets (\ref{dirac}) instead of defining a
set of independent variables in $S^3$, because this would lead us to
very complicated expressions. Therefore, from now on, as we will
work in the configuration space $S^3$ where only Dirac brackets will
be appropriate, the label $D$ (such as it appears in (\ref{dirac}))
will be suppressed.

It is interesting to remark in passing that we can supply a
realization of the variables $x_i$, $p_j$ in terms of canonical
variables and the usual Poisson brackets. Let us consider the
canonical variables $\xi_i, \pi_j$, $i=1,\dots,4$, and the
associated Poisson brackets,

\begin{equation}\label{pb}
\{\phi,\psi\}_P =\sum_i \left( \frac{\partial \phi}{\partial
\pi_i}\frac{\partial \psi}{\partial \xi_i} - \frac{\partial
\psi}{\partial \pi_i}\frac{\partial \phi}{\partial \xi_i} \right) \,
.
\end{equation}
Then, let us define the following relations

\begin{equation}\label{xipi}
x_i({\boldsymbol\xi},{\boldsymbol\pi}) := \frac{\xi_i}{\xi}, \quad
p_i({\boldsymbol\xi},{\boldsymbol\pi}) :=\xi \pi_i - ({\boldsymbol
\pi}{\boldsymbol \xi})\frac{\xi_i}{\xi^2}, \quad\textstyle \xi
=\sqrt{\sum_k (\xi_k)^2} \, ,
\end{equation}
and also Dirac brackets by

\begin{equation}\label{poisson}
\{f({\bf  x},{\bf  p}),g({\bf  x},{\bf  p})\}_D:=
\{f({\boldsymbol\xi},{\boldsymbol\pi}),
g({\boldsymbol\xi},{\boldsymbol\pi})\}_P\,
\end{equation}
where in the last expression the functions depending on
${\boldsymbol\xi},{\boldsymbol\pi}$ have been obtained by the
replacement of ${\bf  x},{\bf  p}$ given in (\ref{xipi}). Then,
(\ref{xipi}), (\ref{poisson}) gives us the desired representation.

From now on, we shall consider Dirac brackets only, so that the
subindex $D$ in brackets like (\ref{dirac}) will be omitted in the
sequel.

The components $J_{ij}$ of the angular momentum introduced in
(\ref{3}) satisfy the following `commutation' relations:
\begin{eqnarray}
 \{J_{ik},J_{lm}\} &=& \delta_{im} J_{kl}+\delta_{kl}J_{im}-\delta_{il}J_{km}-\delta_{km}J_{il} \, ,
  \label{5} \\[1ex]
 \{J_{ik},x_l\} &=& \delta_{lk} x_i-\delta_{il}x_k\, , \label{6}\\[1ex]
  \{J_{ik},p_l\} &=& \delta_{lk} p_i-\delta_{il}p_k\, , \label{6b}
\end{eqnarray}
which coincide with those well known using canonical Poisson
brackets. Hence, the elements $J_{ij}$ are  the generators of the
Lie algebra $\frak{so}(4)$ of the group $SO(4)$. This Lie algebra
has two Casimirs:
\begin{eqnarray}
 H &=& \frac12\;J_{ij}J_{ij}={\bf  p}^2{\bf  x}^2-(\bf { xp})^2
 ={\bf  p}^2\,,  \label{7} \\[1ex]
 C_2 &=& \epsilon_{ijkl}J_{ij}J_{kl}= \bf {xp} = 0\,.   \label{8}
\end{eqnarray}
Thus, the only nontrivial Casimir plays the role of the Hamiltonian
$H$ and it can be seen as the restriction to $S^3$ of the free
Hamiltonian defined in the ambient space. Consequently, the symmetry
algebra is just $\frak{so}(4)$ itself, but taking into account that
this symmetry is realized by a representation in which the second
Casimir (\ref{8}) vanishes.

In the rest of this section we will show that what can be called the
`classical' SGA for this system has the structure of the
$\frak{so}(4,2)$ algebra, including the aforementioned realization
of $\frak{so}(4)$ as a subalgebra.

The generators of the Lie algebra $\frak{so}(4,2)$ will be labeled
$M_{\alpha\beta}$ with $\alpha,\beta=1,\dots,6$. They include the
generators of the symmetry algebra in the form $M_{ij}=J_{ij}$. The
new generators completing the `classical' SGA are $M_{5i}$ and
$M_{6i}$, that behave as vectors with respect to $SO(4)$ and
$M_{56}$ which is an $SO(4)$--scalar. These generators can be
displayed schematically in the form of a $6\times 6$ antisymmetric
matrix as follows:
\begin{equation}\label{9}
M_{\alpha\beta}=\left(
         \begin{array}{cc}
           J_{ij} & \begin{array}{ccc}
                       M_{i5} & M_{i6}
                     \end{array}
            \\[1ex]
\begin{array}{c}
  M_{5i} \\[1ex]
  M_{6i}
\end{array}
            & \begin{array}{cc}
                 0 & M_{56} \\[1ex]
                 M_{65} & 0
               \end{array}
             \\
         \end{array}
       \right)\, ,
\end{equation}
such that $M_{\alpha\beta}=-M_{\beta\alpha}$. The $\frak{so}(4,2)$
`commutation' relations (in the sense of Dirac brackets) are
\begin{equation}\label{10}
  \{M_{\alpha\beta},M_{\gamma\delta}\}=g_{\alpha\delta}M_{\beta\gamma}+
  g_{\beta\gamma}M_{\alpha\delta}-g_{\alpha\gamma}M_{\beta\delta}-g_{\beta\delta}M_{\alpha\gamma}\,,
\end{equation}
where $g_{\alpha\beta}$   is the metric matrix for $SO(4,2)$, given
by the diagonal $6\times 6$ matrix $(1,1,1,1,-1,-1)$. In the
following, we will pay special attention to the commutators of the
generators not belonging to $\frak{so}(4)$:
\begin{eqnarray}
\{M_{i5},M_{56}\}&=&  -M_{i6}\,, \label{11}\\[1ex]
\{M_{i6},M_{56}\}&= &  M_{i5}\,,\label{12}\\[1ex]
\{M_{i5},M_{k6}\}&= &  -\delta_{ik}M_{56}\,,  \label{13}\\[1ex]
\{M_{i5},M_{k5}\}&= &  J_{ik}\,, \label{14}\\[1ex]
\{M_{i6},M_{k6}\}&= &  J_{ik}\label{15} \, .
\end{eqnarray}
Our next objective is to calculate explicit expressions for $M_{5i},
M_{6i},M_{56}$ implementing these commutators. In this process we
need two $SO(4)$ vectors and one $SO(4)$ scalar. From (\ref{6}) and
(\ref{6b}), the vectors at hand are $\bf {x}$ and $\bf {p}$, while
the scalar is just $H$, therefore we can make the following guess
for these generators:
\begin{equation}
M_{i5} =  p_if_5(H),\quad M_{i6} =  x_if_6(H),\quad M_{56} =
f_{56}(H)
\end{equation}
where $f_5,f_6,f_{56}$ are functions to be determined. Remark that
due to the specific realization we have
\begin{equation}
p_i = - J_{ik}x_k \, . \label{pjx}
\end{equation}
Since the Dirac brackets  for $p_i$ in $S^3$ are not familiar (see
(\ref{dirac})), we prefer to use the equivalent expression
(\ref{pjx}), because its factors obey the Dirac brackets with the
familiar form (\ref{dirac})--(\ref{6}). The final solution in this
classical frame can be easily obtained:
\begin{equation}\label{221}
M_{i5} = J_{ik}x_k,\quad M_{i6} = H^{1/2} x_i,\quad M_{56} = H^{1/2}
\, .
\end{equation}

Making use of the  `classical' SGA, we can solve the equation of
motion for the present case. To this end, let us introduce the
functions $A_j^\pm$ as  the following linear combinations of
$M_{5i}$ and $M_{6i}$:
\begin{equation}\label{21}
A_j^\pm:= M_{5j}\mp iM_{6j}\,,\qquad j=1,\dots 4\,.
\end{equation}
The importance of the $SO(4)$ vectors $A_j^\pm$  will have  quite
different implications in  the classical and quantum cases. Clearly,
the equations of motion for $A_j^\pm$ are given by
\begin{equation}\label{22}
\dot A_j^\pm=\{H,A_j^\pm\}=2\sqrt H\{\sqrt H,A_j^\pm\}=\pm 2\sqrt
H\,iA_j^\pm\,,
\end{equation}
where the dot denotes derivative with respect time. Then, (\ref{22})
gives:
\begin{equation}\label{23}
A_j^\pm(t)=\exp\{\pm 2i\,t\sqrt H\}\,A_j^\pm(0)\,,
\end{equation}
which provide the motion $(x_j(t),p_j(t))$. Notice that the
frequency $\omega=2\sqrt H$ depends on $H$ and increases as the
system has higher energies. Another way to express (\ref{23}) is to
say that $A_j^\pm(t)\exp\{\mp 2it\sqrt H\}$ are time-dependent
constants of motion whose values $A_j^\pm(0)$ are fixed from the
initial conditions. Therefore, the symmetries lead to
(time-independent) constants of motion, while the other SGA elements
lead in this way to explicit time-dependent constants of motion
giving the motion.

On the other hand, $A^+_iA^-_i$ is a time-independent $SO(4)$ scalar
function, that   can be computed explicitly:
\begin{equation}
A^+_iA^-_i=A^+_i(0)A^-_i(0)=2H\, , \label{amplitude}
\end{equation}
so that the amplitudes $|A_i^\pm(0)|$ in (\ref{23}) also depend on
the Hamiltonian. We can also compute the value of the quadratic
Casimir of $\frak{so}(4,2)$ with the help of (\ref{amplitude}):
\begin{equation}
M^{\alpha\beta}M_{\alpha\beta}=J_{ij}J_{ij}+2M_{56}^2-2(M_{5i}M_{5i}+M_{6i}M_{6i})
 =2H+2H-2(2H) =0\,. \label{24}
\end{equation}
We see that (\ref{24}) also gives (\ref{amplitude}).

As a final remark, we shall note that some {\it restrictive
relations} among the algebra generators are fulfilled. These
relations will have a crucial role in the construction of the
quantum SGA for the quantum case. In fact, a rather straightforward
calculation show that the tensors given by

\begin{equation}\label{2.27}
T_{ab}:= M_{ac}M_{bd}\,g^{cd}\qquad {\rm and} \qquad
R^{ab}:=\varepsilon^{abcdef}\,M_{cd}M_{ef}\,,
\end{equation}
where $\varepsilon^{abcdef}$ is the complete antisymmetric tensor,
vanish identically, i.e.,

\begin{equation}\label{2.28}
  T_{ab}=0  \qquad {\rm and} \qquad R^{ab}=0\,.
\end{equation}

We shall call relations (\ref{2.28})  {\it restrictive relations}
for the algebra ${\frak so}(4,2)$. These relations are not changed
under the action of the algebra since a direct calculation using
(\ref{10}) gives:

\begin{equation}\label{2.29}
    \{M_{ab},T_{cd}\}=
    g_{ac}\,T_{bd}-g_{bc}\,T_{ad}+g_{ad}\,T_{cb}-g_{bd}\,T_{ca}\,.
\end{equation}
A similar relation holds for $\{M_{ab},R^{cd}\}$.

Here, we conclude the discussion of the classical SGA for $S^3$.

\section{The Quantum Case}

Let us analyze the quantized version of the discussion given in the
previous section. We shall adopt a point of view of starting with
the quantized version of (\ref{10}). In any case and by analogy with
the classical case, we shall work in a realization in which the
Casimirs for the symmetry algebra $\frak{so}(4)$ are the free
Hamiltonian

\begin{equation}\label{25}
H=\frac12\,J_{jk}J_{jk}
\end{equation}
and

\begin{equation}\label{c2q}
 C_2 = \frac12\epsilon_{ijkl}J_{ij}J_{kl} = 0\,.
\end{equation}
We shall present the discussion of the quantum case in the next
following subsections.

\subsection{The symmetry algebra}

The symmetry algebra of $H$  is clearly determined by the operators
$J_{jk}$ that close the algebra $\frak{so}(4)$. We recall (see for
instance \cite{R}) that the algebra $\frak{so}(4)$ is the direct sum
of two copies of $\frak{su}(2)$, i.e.,
$\frak{so}(4)=\frak{su}(2)\oplus \frak{su}(2)$. If we group the six
generators $J_{jk}$ as the components of two vectors  $\bf {
R}=(J_{23},-J_{13},J_{12})$ and $\bf {S}=(J_{14},J_{24},J_{34})$,
then the generators of each copy of $\frak{su}(2)$ are given by $\bf
{ M}=(\bf {R}+\bf { S}$)/2 and $\bf { N}=(\bf { R}-\bf { S})/2$,
respectively. The corresponding $\frak{su}(2)$ Casimirs are $\bf {
M}^2=(\bf { R}+\bf { S})^2/4= (\bf {R}^2+\bf {R}\bf { S}+\bf { S}\bf
{ R}+ \bf { S}^2)/4$, and $\bf { N}^2=(\bf { R}-\bf { S})^2/4=(\bf {
R}^2-\bf { R}\bf { S}-\bf { S}\bf { R}+ \bf { S}^2)/4 $\,. Thus, the
Hamiltonian in (\ref{28}) can be expressed in the form
\begin{equation}
H = 2(\bf { M}^2+\bf { N}^2)
\end{equation}

The Hilbert space spanned by the states with the same energy support
a unitary irreducible  representation (UIR) of $SO(4)$. However, we
must take into account that  in this particular realization we have
that ${\bf RS}+{\bf SR}=C_2 = \frac12\epsilon_{ijkl}J_{ij}J_{kl}=0$,
so that the value for both of the Casimirs for  $\frak{su}(2)$
 coincides: $\bf { M}^2=\bf { N}^2 =j(j+1)$
with $j=0,1/2,1,3/2,\dots$. Then, each representation supported by
states with the same energy is symmetric and hence the spectrum of
the Hamiltonian is given by
\begin{equation}\label{espectrum}
{\rm Spec}(H) = 4j(j+1)= n(n+2),\qquad n=0,1,\dots
\end{equation}
The  $n$-th energy level has the value $j=n/2$ for each of the
$\frak{su}(2)$--components, and consequently the degeneracy of this
energy level is $(n+1)^2$.

\subsection{The quantum Spectrum Generating Algebra.}

To begin with, let us take the quantized version of formula
(\ref{10}). This quantized formula gives us the relation between the
generators of $SO(4,2)$, where these generators are operators on a
certain Hilbert space and the Dirac brackets have been replaced by
commutators. It reads:

\begin{equation}\label{3.1}
 [M_{ab},M_{cd}]=-i(g_{ad}M_{bc}+g_{bc}M_{ad}-g_{ac}M_{bd}-g_{bd}M_{ac})\,,
\end{equation}
with $a,b=1,\dots,6$ and $\{g_{ab}\}$ is a $6\times 6$ diagonal
matrix with diagonal $(1,1,1,1,-1,-1)$. Take the indices
$i,j=1,2,3,4$ and introduce the following notation:

\begin{equation}\label{3.2}
M_{ij}=J_{ij},\quad M_{i5}=K_{i},\quad M_{i6}=L_{i},\quad
M_{56}=h\,.
\end{equation}
We use this notation (\ref{3.2}) in formula (\ref{3.1}) so as to
obtain

\begin{eqnarray}
&[J_{ik},J_{lm}]=-i(\delta_{im}J_{kl}+\delta_{kl}J_{im}-\delta_{il}J_{km}-\delta_{km}J_{il})\nonumber\\[2ex]
&[J_{ik},K_l]=-i(\delta_{kl}K_i-\delta_{il}K_k),\quad
[J_{ik},L_l]=-i(\delta_{kl}L_i-\delta_{il}L_k)\nonumber\\[2ex]
&[K_i,K_j]=[L_i,L_j]=-iJ_{ij},\quad
[K_i,L_j]=i\delta_{ij}h\nonumber\\[2ex]
&[K_i,h]=iL_i,\quad[L_i,h]=-iK_i\,.\label{3.3}
\end{eqnarray}

One of the important features of the SGA are the ladder operators
which will be used in order to construct the Hilbert space of pure
states of the system. By close analogy with the classical case (see
equation (\ref{21})), we define

\begin{equation}\label{3.4}
 A^{\pm}_i=K_i\mp iL_i\,,\qquad i=1,2,3,4\,.
\end{equation}
These operators behave as vectors with respect to the generators
$J_{ij}$ of the algebra $\frak{so}(4)$, as shown in the second row
of (\ref{3.3}). They satisfy the following commutation relations:

\begin{eqnarray}
&[A^+_i,A^-_j]=-2iJ_{ij}-2\delta_{ij}h,\quad[A^+_i,A^+_j]
=[A^-_i,A^-_j]=0,\nonumber\\[2ex]
&hA^+_i=A^+_i(h+1),\quad hA^-_i=A^-_i(h-1)\,.\label{3.5}
\end{eqnarray}

Now, our purpose is to identify the quantum system with spectrum
generating algebra given by $SO(4,2)$. In order to accomplish this,
we need to find some restrictive relations for the generators of the
algebra $\frak{so}(4,2)$. This reflect the fact that, in general,
the Hilbert space obtained with this technique is not the Hilbert
space which supports a general representation of $SO(4,2)$. This
kind of restrictive relations is very well known, although not very
well identified often.

In order to understand this fact, let us consider the simple example
of $SO(3)$. Its generators satisfy the familiar commutation
relations

\begin{equation}\label{3.6}
[S_i,S_j]=i\epsilon_{ijk}S_k\,.
\end{equation}
The representations of $SO(3)$ are well known and are labelled by
the eigenvalues $s$ of the Casimir operator. At the same time, we
can impose on the generators $S_i$ some additional conditions. For
instance:

\begin{equation}\label{3.7}
T_{ij}=S_i S_j+S_j S_i-\frac{1}{2}\,\delta_{ij}=0\,,
\end{equation}
which are compatible with (\ref{3.6}). In fact,

\begin{equation}\label{3.8}
[S_l,T_{ij}]=i(\epsilon_{lik}T_{kj}+\epsilon_{lij}T_{ik})\,.
\end{equation}

Condition (\ref{3.7}) is a restrictive relation of the mentioned
type.  Then, if in the space supporting all representations of
$SO(3)$, we are to define a subspace $V$ satisfying the condition
$T_{ij}\psi=0$, the action of the generators of $SO(3)$ on $V$
should not leave $V$. In the chosen example, it becomes clear that
the space in which this condition is satisfied corresponds to the
choice of $s=1/2$. For higher values of the spin, the restrictive
relations for $SO(3)$ involve higher powers of $S_i$.

In the case of a non compact algebra like $\frak{so}(4,2)$, the
situation is more complicated so that even a quadratic restriction
relation may define a subspace $V$ of infinite dimension. Thus, the
first problem that we have to solve is to find a general expression
for the restrictive relations concerning $SO(4,2)$. These
restrictive relations will be given by operators (as in (\ref{3.7})
such that their commutators with the generators $M_{ab}$ are linear
on these operators. This task is not difficult if we construct the
operators providing the restrictive relations as tensors constructed
from $M_{ab}$.

One more remark is in order here. As we have seen in the classical
case, all generators of $SO(4,2)$ (6+4+4+1=15), were build by using
$x_i$ and $p_i$. This shows the existence of some relations between
the $M_{ab}$. In the quantum case, these relations cannot be valid
in the general representation of $SO(4,2)$, but they may hold for
particular representations and this is the case here.

To produce this construction, let us go back to tensors
(\ref{2.27}). In our representation we have changed the meaning of
the objects $M_{ab}$ that do not represent functions any more like
in (\ref{2.27}), but Hermitian operators instead. Since these
operators do not commute, we have to symmetrize (\ref{2.27}) so that
the new expression for $T_{ab}$ should be given by

\begin{equation}\label{3.9}
T_{ab}=(M_{ad}M_{be}+M_{be}M_{ad})g^{de}\,,
\end{equation}
which is covariant with respect to the generators of $SO(4,2)$:

\begin{equation}\label{3.10}
[M_{ab},T_{cd}]=i(g_{ac}T_{bd}-g_{bc}T_{ad}+g_{ad}T_{cb}-g_{bd}T_{ca})\,.
\end{equation}
Needless to say that (\ref{3.10}) replaces (\ref{2.29}) in the
present discussion. Moreover, if we add a constant term $cg_{ab}$,
where $c$ is arbitrary, to $T_{ab}$:

\begin{equation}\label{3.11}
\widetilde{T}_{ab}=T_{ab}+cg_{ab}\,,
\end{equation}
then, $\widetilde T_{ab}$ will satisfy the same equation.

A second restrictive relation is given by the following tensor,
which is clearly the quantized version of the second tensor in
(\ref{2.27}):

\begin{equation}\label{3.11}
R^{ab}=\epsilon^{abcdef}(M_{cd}M_{ef}+M_{ef}M_{cd})\,.
\end{equation}

Then, we need to express the components of $\widetilde T_{ab}$ and
$R_{ab}$ in terms of the components $M_{ab}$. We shall do it using
the notation introduced in (\ref{3.2}). The components of $T_{ab}$
are

\begin{eqnarray}
&\widetilde{T}_{ij}= J_{ik}J_{jk}+J_{jk}J_{ik}-(K_i K_j+K_j K_i+L_i
L_j+L_j L_i)+c
g_{ij}\,,\label{3.13}\\[2ex]
&\widetilde{T}_{5i}= -(h L_i+L_i
h)-(J_{ij}K_j+K_jJ_{ij})\,,\label{3.14}\\[2ex]
&\widetilde{T}_{6i}= h K_i+K_i
h-(J_{ij}L_j+L_jJ_{ij})\,,\label{3.15}\\[2ex]
&\widetilde{T}_{56}= K_i L_i+L_i K_i\,,\label{3.16}\\[2ex]
&\widetilde{T}_{55}= 2(K_i^2-h^2)-c\,,\label{3.17}\\[2ex]
&\widetilde{T}_{66}=2( L_i^2-h^2)-c\,.\label{3.18}
\end{eqnarray}
 Also note that we are using the
convention of sum over repeated indices.  This is also true in
(\ref{3.17}) and (\ref{3.18}), where we have the terms
$L_i^2=L_iL_i=\sum_{i=1}^4 L_i^2$ and also
$K_i^2=K_iK_i=\sum_{i=1}^4 K_i^2$. We shall use this convention from
now on. For instance, $X_i^2$ will denote $\sum_{i=1}^n X_i^2$, etc.

The components of $R^{ab}$ are

\begin{eqnarray}
&R^{ij}= K_i L_j+L_j K_i-(L_i K_j+K_j L_i)-2h
J_{ij}\,,\label{3.19}\\[2ex]
&R^{5i}=
\epsilon_{ijkl}(L_jJ_{kl}+J_{kl}L_j)\,,\label{3.20}\\[2ex]
&R^{6i}=
\epsilon_{ijkl}(K_jJ_{kl}+J_{kl}K_j)\,,\label{3.21}\\[2ex]
&R^{56}= \epsilon_{ijkl}J_{ij}J_{kl}\,.\label{3.22}
\end{eqnarray}

Once we have defined the operators giving the restriction relations,
we can write these relations as:

\begin{equation}\label{3.23}
    \widetilde{T}_{ab}=0, \qquad R^{ab}=0\,.
\end{equation}

The next step is finding operators that can play the role of
position operators. Thus, we find four operators $X_i$, $i=1,2,3,4$
subject to these conditions: i.) the operators $X_i$ commute with
each other and ii.) $ X_i^2=1$, i.e., they determine the position on
the sphere $S^3$. We have the following Ansatz for the $X_i$:

\begin{equation}\label{3.24}
X_i=f(h)A^+_i+g(h)A^-_i\,.
\end{equation}
This Ansatz is motivated by the fact that $X_i$ should behave as a
vector with respect to the representations of $SO(4)$. Vectors with
respect these representations are those with components $L_i$ and
$K_i$ only. These vectors are linear combinations of $A_i^\pm$. The
most general linear combination of these operators is given by
(\ref{3.24}) since $h$ behaves like an scalar. We know that from the
restrictive relations, we can express any scalar, e.g., $L^2_i$,
$K_i^2$, etc via $h$.

After this definition, in order to obtain the commutation relations
for the $X_i$, we shall use (\ref{3.5}). This gives:

\begin{eqnarray}\label{3.25}
&[X_i,X_j]=-4iJ_{ij}f(h)g(h)+f(h)(g(h-1)-g(h))(A^+_i A^-_j-A^+_j
A^-_i)\nonumber\\[2ex]
&+g(h)(f(h+1)-f(h))(A^-_iA^+_j-A^-_j A^+_i)\,.
\end{eqnarray}

Condition $R^{ab}=0$ implies the annihilation of $R^{ij}$ in
(\ref{3.19}). From this fact and definition (\ref{3.4}), we get

\begin{eqnarray}
&(A^+_i A^-_j-A^+_j A^-_i)=2i(h-1)J_{ij}\,,\nonumber\\[2ex]
& (A^-_iA^+_j-A^-_j A^+_i)=-2i(h+1)J_{ij}\,.\label{3.26}
\end{eqnarray}
Then, we can carry (\ref{3.26}) into (\ref{3.25}) to obtain the
commutation relations between the $X_i$:

\begin{equation}\label{3.27}
[X_i,X_j]=2i J_{ij}((h-1)f(h)g(h-1)-(h+1)g(h)f(h+1))\,.
\end{equation}
However, we have imposed the condition that these commutators must
vanish. This condition is obviously satisfied if

\begin{equation}\label{3.28}
(h-1)f(h)g(h-1)-(h+1)g(h)f(h+1)=0\,,
\end{equation}
a finite difference equation that we have to solve.

In addition, operators $X_i$ must be Hermitian. Since the operators
$A_i^\pm$ are adjoint of each other, definition (\ref{3.24}) and the
second row of (\ref{3.5}), the Hermiticity property for $X_i$
implies a second condition on the functions $f(h)$ and $g(h)$, which
is

\begin{equation}\label{3.29}
g(h)=f^*(h+1)\,,
\end{equation}
where the asterisk denotes complex conjugation. Using (\ref{3.29})
into (\ref{3.28}), we have

\begin{equation}\label{3.30}
(h-1)f(h)f^*(h)-(h+1)f(h+1)f^*(h+1)=0\,.
\end{equation}

The solution of (\ref{3.30}) is given by

\begin{equation}\label{3.31}
f(h)=\frac{C}{\sqrt{h(h-1)}}\,,
\end{equation}
where $C$ is an arbitrary constant that may be chosen to be real.
Using (\ref{3.31}) and (\ref{3.29}) in (\ref{3.24}), we get

\begin{eqnarray}\label{3.32}
X_i=\frac{C}{\sqrt{h(h-1)}}A^+_i+\frac{C}{\sqrt{h(h+1)}}A^-_i
=C\frac{1}{\sqrt{h}}(A^+_i+A^-_i)\frac{1}{\sqrt{h}}\,.
\end{eqnarray}
We have used the second row of (\ref{3.5}) to prove the second
identity in (\ref{3.32}).

The next step is to calculate the sum $ X_i^2$. This gives:

\begin{eqnarray}
X_i^2=C^2\frac{1}{\sqrt{h}}\left\{(A^+_i+A^-_i)\frac{1}{h}(A^+_i+A^-_i)\right\}\frac{1}{\sqrt{h}}\nonumber\\[2ex]
=C^2\frac{1}{\sqrt{h}}\left\{(A^+_i+A^-_i)(A^+_i\frac{1}{h+1}+A^-_i\frac{1}{h-1})\right\}
\frac{1}{\sqrt{h}}\,.\label{3.33}
\end{eqnarray}

Then, we recall that the restrictive relations mean that all
components of $\widetilde T$ and $R$ are equal to zero. In
particular, if we equate to zero (\ref{3.16}-\ref{3.18}) and use
definition (\ref{3.4}), we conclude that

\begin{equation}\label{3.34}
(A^+_i)^2=(A^-_i)^2=0\,.
\end{equation}
Also, using (\ref{3.17}), (\ref{3.18}) and one of the commutation
relations (\ref{3.3}), we obtain

\begin{eqnarray}
&A^+_iA^-_i=K_i^2+L_i^2-4h=2h^2+c-4h\nonumber\\[2ex]
&A^-_iA^+_i=K_i^2+L_i^2+4h=2h^2+c+4h\,.\label{3.35}
\end{eqnarray}
Then, using (\ref{3.34}) and (\ref{3.35}) into (\ref{3.33}), we
obtain

\begin{equation}\label{3.36}
X_i^2=C^2\frac{1}{\sqrt{h}}\,\left(\frac{2h^2+c+4h}{h+1}+\frac{2h^2+c-4h}{h-1}\right)\,\frac{1}{\sqrt{h}}\,.
\end{equation}
The choice $c=2$ gives

\begin{equation}\label{3.37}
X_i^2=4C^2\,.
\end{equation}

Since $X_i^2=1$, this implies that $C=1/2$ and therefore the
expression for $X_i$ should be

\begin{equation}\label{3.38}
X_i=\frac{1}{2}\frac{1}{\sqrt{h}}\,\left(A^+_i+A^-_i\right)\,\frac{1}{\sqrt{h}}\,.
\end{equation}

The choice on the constant $c$ is also important in establishing a
relation between the Casimir of $SO(4)$ and $h$. If we calculate the
trace of the $4\times 4$ matrix $\widetilde T_{ij}$ and take into
account that all entries of this matrix vanish so that this trace
must also vanish, we obtain from (\ref{3.13})

\begin{equation}\label{3.39}
\frac{1}{2}J_{ik}J_{ik}=\frac{1}{2}(K_i^2+L_i^2)-c\,.
\end{equation}
Taking into account that (\ref{3.17}) and (\ref{3.18}) also vanish,
we have that

\begin{equation}\label{3.40}
K_i^2+L_i^2=2h^2+c\,,
\end{equation}
which with (\ref{3.39}) gives

\begin{equation}\label{3.41}
\frac{1}{2}J_{ik}J_{ik}=h^2-\frac{c}{2}\,.
\end{equation}
With the choice $c=2$, (\ref{3.41}) becomes

\begin{equation}\label{3.42}
\frac{1}{2}J_{ik}J_{ik}=h^2-1\,.
\end{equation}

From the theory of representation of Lie groups \cite{R}, we know
that if the second Casimir  is $\epsilon_{ijkl}J_{ij}J_{kl}=0$, then
the first Casimir can be represented via a positive operator
$\gamma$ as

\begin{equation}\label{3.43}
\frac{1}{2}J_{ik}J_{ik}=\gamma(\gamma+2)\,,
\end{equation}
which gives the following expression for $h$

\begin{equation}\label{3.44}
h=\gamma+1\,.
\end{equation}

In conclusion, we have constructed on the algebra $\frak{so}(4,2)$ a
set of restrictive relations, which defines a subspace of the space
supporting the representations of the algebra. We have found that,
on this subspace, the operators $X_i$, $i=1,2,3,4$ act. These
operators are position operators on a configuration space which is
the homogeneous space for the symmetry algebra $SO(4)$. Moreover,
taking into account that the expressions (\ref{3.13}-\ref{3.18})
vanish, we can express the generators of $SO(4,2)$ in terms of the
operators $J_{ij}$,$X_i$ and $h$. It remains to prove this latter
statement. Let us do it for $L_i$, for example. From the vanishing
of (\ref{3.14}), we have

\begin{equation}\label{3.45}
hL_i+L_i h=-(J_{ij}K_j+K_jJ_{ij})\,.
\end{equation}
Using (\ref{3.4}), the left hand side of this equation becomes

\begin{eqnarray}
hL_i+L_i h=ih
\frac{A^+_i-A^-_i}{2}+i\frac{A^+_i-A^-_i}{2}h\nonumber\\[2ex]
=if(h)\frac{A^+_i-A^-_i}{2}f(h)=f(h)L_if(h)\,.\label{3.46}
\end{eqnarray}

It remains to determine the function $f(h)$ in (\ref{3.46}). This
function has to satisfy

\begin{equation}\label{3.47}
h(A^+_i-A^-_i)+(A^+_i-A^-_i)h=f(h)(A^+_i-A^-_i)f(h)\,.
\end{equation}
With the help of equations (\ref{3.5}), we can rewrite (\ref{3.47})
in the following form

\begin{equation}\label{3.48}
A^+_i(2h+1)-A^-_i(2h-1)=A^+_if(h)f(h+1)-A^-_if(h)f(h-1)\,,
\end{equation}
from which we can easily derive a finite difference equation for the
function $f(h)$:

\begin{equation}\label{3.49}
f(h)f(h+1)=2h+1\,,
\end{equation}
which has the following solution

\begin{equation}\label{3.50}
f(h)=2\frac{\Gamma(\frac{h}{2}+\frac{3}{4})}{\Gamma(\frac{h}{2}+\frac{1}{4})}\,,
\end{equation}
where $\Gamma(x)$ is the Euler function. Indeed,

\begin{equation}\label{3.51}
f(h)f(h+1)=4\frac{\Gamma(\frac{h}{2}+\frac{3}{4})}{\Gamma(\frac{h}{2}+\frac{1}{4})}
\frac{\Gamma(\frac{h}{2}+\frac{5}{4})}{\Gamma(\frac{h}{2}+\frac{3}{4})}
=\frac{\Gamma(\frac{h}{2}+\frac{5}{4})}{\Gamma(\frac{h}{2}+\frac{1}{4})}=4(\frac{h}{2}+\frac{1}{4})=2h+1\,.
\end{equation}

Once we have obtained $f(h)$, we can rewrite (\ref{3.45}) in the
following form

\begin{equation}\label{3.52}
    f(h)L_if(h)=-(J_{ij}K_j+K_jJ_{ij})=-\sqrt{h}(J_{ij}X_j+X_jJ_{ij})\sqrt{h}\,,
\end{equation}
where in the last identity, we have used (\ref{3.38}). Thus, the
final result for $L_i$ is given by

\begin{equation}\label{3.53}
L_i=-(f(h))^{-1}\sqrt{h}(J_{ij}X_j+X_jJ_{ij})\sqrt{h}(f(h))^{-1}\,.
\end{equation}
This is the form of the generators $M_{i6}$ of the algebra
$\frak{so}(4,2)$. For $M_{i5}=K_i$ we note that (\ref{3.4}) gives
$K_i=1/2(A_i^++A_i^-)$. Then, (\ref{3.38}) gives

\begin{equation}\label{3.54}
 M_{i5}=K_i=\sqrt h X_i\sqrt h\,.
\end{equation}
The others are

\begin{equation}\label{3.55}
 M_{ij}=J_{ij}\,,\qquad M_{56}=\gamma+1=h\,.
\end{equation}
This concludes the construction of the generators of the quantum
spectrum generating algebra.

{\bf Remark}.- This representation has three Casimirs, which are the
following:

\begin{equation}\label{3.56}
C_2=M_{ab}M^{ab}=\frac 12\,(\widetilde{T}_{ab}-cg_{ab})g^{ab}=-\frac
12\,cg_{ab}g^{ab}= -3c=-6\,,
\end{equation}

\begin{equation}\label{3.57}
\widetilde C_2=R^{ab}g_{ab}=0
\end{equation}
and

\begin{equation}\label{3.58}
C_3=M_{ab}M^{bc}M_c^a=T_{ab}M^{ab}=(\widetilde{T}_{ab}-cg_{ab})M^{ab}=
-cg_{ab}M^{ab}=0\,.
\end{equation}

\subsection{Ladder representation}

We have already mentioned that the spectrum of the operator $\gamma$
is the set of nonegative integers $n=0,1,2,\dots$ and consequently
the spectrum of the Hamiltonian $H$ is given by $n(n+2)$. These
energy levels can be connected by a ladder representation, which
follows from the `quantization' of the functions (\ref{21}) leading
to the operators $A_i^\pm$.   In fact, after the identities of the
last row in (\ref{3.3}), we have

\begin{eqnarray}
[\gamma,A_i^\pm] &=& \pm A_i^\pm \label{62}
\\[1.5ex]
[A_i^-,A_i^+]  &=& 2(\gamma+1)\,. \label{63}
\end{eqnarray}
Now, let us assume that there exists an eigenvector $|n\rangle$ of
$\gamma$ with eigenvalue $n$, i.e., $\gamma|n\rangle=n|n\rangle$.
Then, (\ref{62}) yields to
\begin{equation}\label{65}
\gamma A_i^\pm|n\rangle=(n\pm 1)A_i^\pm|n\rangle\,,
\end{equation}
so that $A_i^\pm|n\rangle$ is an eigenvector of $\gamma$, provided
that it does not vanish, with eigenvalue $n\pm1$.

The ground state $|0\rangle$ is defined by
\begin{equation}
A_i^-|0\rangle=0\,,\qquad i=1,2,3,4\,.
\end{equation}
The only function satisfying these conditions is the constant
function $\psi_0$ which is normalizable on $S^3$ and represents the
ground state of $H$ as well as the lowest weight vector of the
$\frak{so}(4,2)$ UIR. The other eigenstates of $H$ can be obtained
from $\psi_0$ by applying the raising operators $A_i^+$, which
connects each energy level with the next with higher energy and
$J_{ij}$, which connects states with the same energy. These
eigenstates can be written as

\begin{equation}\label{eigen}
  \psi_{\mu_1\mu_2\dots\mu_n}({\bf
  x})=A_{\mu_1}^+\,A_{\mu_2}^+\,\dots \,A_{\mu_n}^+\,\psi_0({\bf
  x})\,.
\end{equation}
Note that after the relations $[A_i^+,A_j^+]=0$ in (\ref{3.5}) and
$(A_i^+)^2=0$ in (\ref{3.34}), it is obvious that
$\psi_{\mu_1\mu_2\dots\mu_n}({\bf  x})$ is symmetric under the
interchange of the subindices $\mu_i$ and that products of the form
$\psi_{\mu_1\dots \mu_i\dots\mu_j\dots\mu_n}({\bf
x})\,g^{\mu_i\mu_j}$ vanish. Functions of this kind are called {\it
harmonic polynomials}.

Some additional relations obtained with the help of the ladder
operators $A_i^+$ and $A_i^-$ can be obtained. For instance (we
recall that summation over repeated indices still applies),
\begin{equation}\label{67}
    -A_i^-A_i^+= M_{5i}^2+M_{6i}^2-i[M_{5i},M_{6i}]\,.
\end{equation}
We know that the bracket in (\ref{67}) is $iM_{56}=i(\gamma+1)$.
From the restrictive relations (\ref{3.17}) and (\ref{3.18}), we
obtain the following results:

\begin{eqnarray}
 M_{5i}^2  = \frac14\; K_i^2 = \frac14\;\{ \gamma(\gamma+2)+2\} \label{68} \\[2ex]
M_{6i}^2 = \frac14\; L_i^2 = \frac14\;\{ (\gamma+1)^2+2\}\,.
\label{69}
\end{eqnarray}

Remark again that all levels save for the ground state are
degenerate, due to the symmetry algebra. The dimension of the
$n$-level being $(n+1)^2$, $n = 0,1\dots$. An schematic picture can
be  seen in Figure 1.

\begin{figure}
\includegraphics[width=0.8\textwidth]{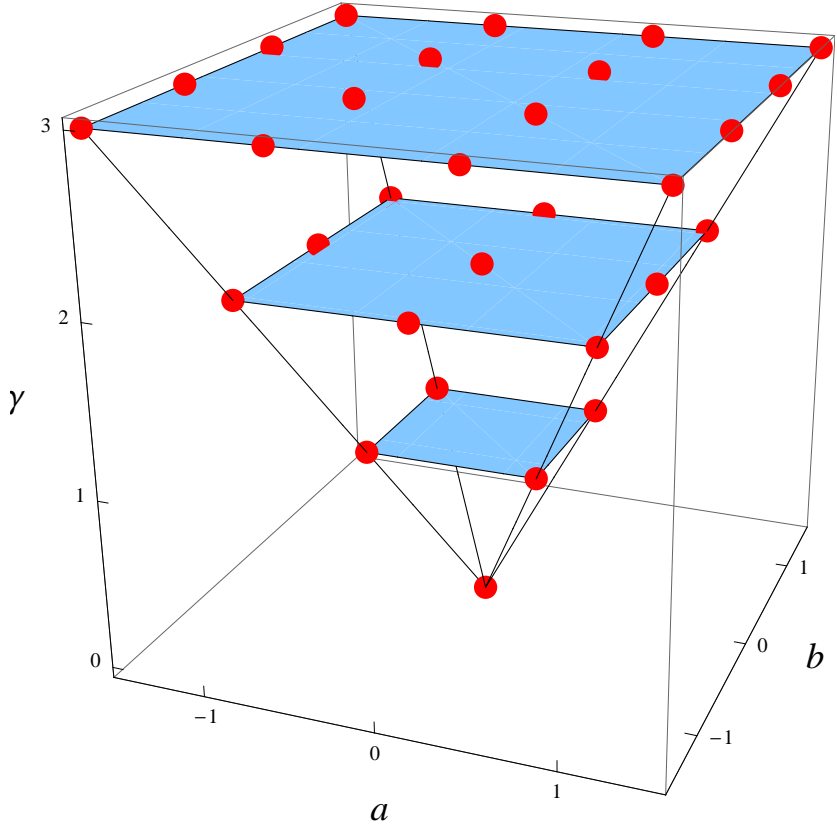}
\caption{Degeneracy levels.}
  \label{f1}
\end{figure}

\section{Another approach to the quantum case}

Another equivalent way in order to find the SGA for the free
particle in $S^3$ comes naturally by considering the canonical
quantization of the Dirac brackets in (\ref{dirac}). In this
context, the canonical variables in (\ref{dirac}) become densely
defined  Hermitian operators on $L^2(S^3)$ and the Dirac brackets
are transformed into commutators multiplied by $i$ whenever
necessary for Hermiticity reasons. To simplify the notation, a
quantum observable will be denoted by the capital letter that
denotes the corresponding classical observable, so that the quantum
Hermitian operators corresponding to $x_i$, $p_i$ will be designed
by $X_i,P_i$, respectively. They fulfill the commutation relations,
which result from the canonical quantization of the above Dirac
deformed Poisson brackets obtained by replacing the brackets
$\{a,b\}=c$ by the commutators $[A,B]=-i\,C$ (where $A,B,C$ denote
the quantum operators of the classical analogs $a,b,c$). In this
case (\ref{dirac}) becomes:

\begin{equation}\label{4.1}
 [X_j,X_k]=0\,,\quad    [P_j,X_k]=-i(\delta_{jk}-X_jX_k)\,, \quad
[P_j,P_k]=-i J_{jk}\,.
\end{equation}
The operators $J_{jk}$ keep the same expression as in the classical
case taking care of the ordering,

\begin{equation}\label{4.2}
J_{jk} = X_j P_k -X_k P_j \, .
\end{equation}
From (\ref{4.2}), we see that these operators are also Hermitian and
$J_{jk} = - J_{kj}$. The above commutators imply that the operators
$J_{jk}$ close indeed  the Lie algebra $\frak {so}(4)$  and that
${\bf  X}$ and ${\bf  P}$ are four vectors under the $J_{jk}$
generators, i.e., relations (\ref{5})-(\ref{6b}) are implemented
through commutators at the quantum level.

In the classical discussion, we have established the need of a
constraint (\ref{3}) and a gauge fixing condition (\ref{2.5}) for
the classical variables, and the same has to be done in the quantum
frame for the corresponding quantum operators.
 Here, the gauge condition is
chosen to be
\begin{equation}\label{4.3}
{\bf  X}^2 = 1 \,,
\end{equation}
which is consistent with the fact that the configuration space is
$S^3$. The classical constrain $\bf { x  p}=0$ should be here
expressed in terms of the symmetrized operator $(\bf { X P} +\bf { P
X})/2$. In this respect, first note that that the generators $\bf
{XP}$ and $\bf {PX}$ commute with any position $X_i$ or momentum
$P_j$ operators. Then, from (\ref{4.1}) and (\ref{4.2}), we obtain
the following relations:

\begin{eqnarray}
\bf { XP}=\bf { PX}+3i   \label{4.4}
 \\[1.5ex]
J_{ik}X_k=-P_i+X_i(\bf { PX})  \label{4.5}\\[1.5ex]
X_kJ_{ik}= -P_i+X_i(\bf { XP})\,. \label{4.6}
\end{eqnarray}
Summing (\ref{4.5}) and (\ref{4.6}), we have

\begin{equation}\label{4.7}
\frac 12\, (J_{ik}X_k+X_kJ_{ik})= -P_i+\frac 12\,X_i(\bf { PX}+ \bf
{ XP})\,.
\end{equation}
We must be aware that the initial algebra  (\ref{4.1}) generated by
$\{\bf {X},\bf {P}\}$ remains invariant if we replace the `momentum'
operators by another set $\bf {P} \to \bf {P} + \alpha \bf {X}$,
$\alpha$ being a number or a central element. We can use this
freedom to get simpler expressions. Thus, we define a new momentum
in the form

\begin{equation}\label{4.8}
\widetilde{\bf { P}}:=\bf { P}- \frac 12\,X_i(\bf { PX}+\bf {
XP})\,.
\end{equation}
We can interpret $\widetilde{\bf { P}}$ as the quantum analog of the
projection of the vector $\bf { P}$ on the tangent plane to the
point of the sphere characterized by the vector $\bf { X}$. Thus,
$\widetilde{\bf { P}}$ should be the proper definition for the
quantum momentum operator corresponding to the motion on $S^3$.
Since $\widetilde {J_{ik}}:=X_i\widetilde{P_k}-
 X_k\widetilde{P_i}=X_iP_k-P_kX_i=J_{ik}$,  the set $\{\bf {X},\widetilde{\bf {P}}\}$ has the same formal
commutation relations as (\ref{4.1}). This leads to the following
simplified relations:

\begin{eqnarray}
\bf { X}\widetilde{\bf { P}}+ \widetilde{\bf { P}}\bf { X}=0
\,, \label{4.9}\\[2ex]
 \bf { X}\widetilde{\bf {P}}=-\widetilde{\bf { P}}\bf {X}=\frac{3i}{2}\,,
 \label{4.10}\\[2ex]
 \frac 12\, (J_{ik}X_k+X_kJ_{ik})= -\widetilde{P}_i\,.
 \label{4.11}
\end{eqnarray}
Relations (\ref{4.3}), (\ref{4.9}) and (\ref{4.11}) are the quantum
versions  of the classical formulas (\ref{2.5}), (\ref{3}) and
(\ref{pjx}), respectively. In the sequel, we shall always use
$\widetilde P_i$ instead of $P_i$, although henceforth we shall
remove the tilde for convenience.

As a remark, let us say that we can provide a realization of the
operators $X_i,P_j$, satisfying the previous relations,  in terms of
the canonical operators in $\mathbb R^4$ (which are uniquely defined
up to a unitary equivalence). This realization is given by
\begin{equation}\label{4.12}
\begin{array}{l}
X_i({\boldsymbol\Xi},{\boldsymbol\Pi}) := \frac{\Xi_i}{\Xi}\,,
\\[2ex]
P_i({\boldsymbol\Xi},{\boldsymbol\Pi}) := \frac12(\Xi \Pi_i +\Pi_i
\Xi) - \left( ({\boldsymbol \Pi}{\boldsymbol \Xi}+ {\boldsymbol
\Xi}{\boldsymbol \Pi})\frac{\Xi_i}{4\,\Xi^2} +
\frac{\Xi_i}{4\,\Xi^2}({\boldsymbol \Pi}{\boldsymbol \Xi}+
{\boldsymbol \Xi}{\boldsymbol \Pi}) \right)\,,
\end{array}
\end{equation}
where ${\boldsymbol \Pi}$ and ${\boldsymbol \Xi}$ satisfy the
canonical commutators
\begin{equation}\label{4.13}
[\Xi_j,\Pi_k] = i\,\delta_{ij},\quad [\Xi_j,\Xi_k] =[\Pi_j,\Pi_k] =0
\end{equation}
and $ \Xi =\sqrt{\sum_k (\Xi_k)^2}$. This is the quantum analog of
the classical expressions given in (\ref{xipi}).

Likewise the classical case, one of the $\frak{so}(4)$ Casimirs
plays the role of the Hamiltonian operator
\begin{equation}\label{4.14}
    H=\frac12\;J_{jk}J_{jk}={\bf  P}^2-(\bf { PX})(\bf {XP})={\bf  P}^2-9/4\,,
\end{equation}
where we have taken into account the restrictions on $\bf {X}$ and
$\bf {P}$ of (\ref{4.3}), (\ref{4.9}) and (\ref{4.10}). In the
$\frak{so}(4)$ realization given by (\ref{4.1}) and (\ref{4.2}), the
second Casimir vanishes:

\begin{equation}\label{4.15}
 C_2 = \frac12\epsilon^{ijkl}J_{ij}J_{kl} = 0\,.
\end{equation}

\subsection{The quantum Spectrum Generating Algebra}

In the context of the point of view of the present section,  in
order to construct the Spectrum Generating Algebra (SGA) for the
free particle $S^3$  we will consider the algebra generated by all
the operators $\{X_i,P_j,J_{kl},H\}$, including the Hamiltonian.

We must compute the commutation of $X_i$ and $P_j$ with $H$. From
(\ref{4.1})-(\ref{4.2}), after some straightforward computations, we
get
%
\begin{equation}\label{4.17}
[H,X_i]=-i\,2 P_i
\end{equation}
and

\begin{equation}\label{4.21}
[H,P_i]=-i\bf g((2H+3/2)X_i -2P_i\bf g) \,.
\end{equation}
%
%
The commutators (\ref{4.17}) and (\ref{4.21}) give a linear action
of $H$ on the elements having the form $\alpha(H)X_i + \beta(H)
P_i$, where the coefficients $\alpha,\beta$ may depend on $H$. We
can diagonalize this action as an eigenvalue problem:

\begin{equation}\label{4.23}
[H, V_i^\pm] = \lambda^\pm(H) V_i^\pm\,.
\end{equation}
The solution of this eigenvalue equation is straightforward. The
respective eigenvalues and eigenvectors are
\begin{eqnarray}
\lambda^\pm(H) =-1\pm 2h\,,\qquad {\rm with}\qquad h:=\sqrt{H+1}\,,
\label{4.24}\\[2ex]
V_i^\pm = -i (\pm h +1/2) X_i - P_i \, ,\label{4.25}
\end{eqnarray}
where we must taking care of the order of  $h$ and $X_i$. The
eigenvectors $V_i^\pm$ are defined up to global factors that may
depend on $h$, this freedom will be used later.

Let us go back to the discussion on Section 3.1. There, we have
shown that the operator $h$, as defined in (\ref{4.24}) has a purely
discrete spectrum coinciding with the set of natural numbers. We can
summarize this in the following formulas
\begin{equation}\label{4.26}
H = h^2-1,\quad {\rm Spec}(h) = 1,2,\dots\,.
\end{equation}
Now, let us rewrite (\ref{4.23}) as

\begin{equation}\label{4.27}
 (H-\lambda^\pm)V_i^\pm=V_i^\pm H\,.
\end{equation}
Taking into account (\ref{4.26}) and replacing $\lambda^\pm$ into
(\ref{4.27}), we have
%
\begin{equation}\label{4.29}
(h\mp 1)V_i^\pm=V_i^\pm h\,.
\end{equation}
Therefore,  $V_i^\pm$ act as lowering and raising operators for $h$,
changing its eigenvalues in one unit.

Our next task is to express the initial (quadratic) algebra
generated by $ \{ X_i,P_j, H, J_{ik}\} $ in terms of the most
appropriate  basis $ \{ V_i^\pm, h, J_{ik}\} $. In order to
calculate the commutation of $V_i^\pm$ among themselves, we must
express $X_i$ and  $P_i$ in terms of these eigenvectors (or
eigenoperators) $V_i^\pm$:
\begin{equation}\label{4.30}
 X_i=
 \frac{i}{2h}\;\big(V_i^+-V_i^-\big)\,,\qquad
 P_i=- \frac{1}{2h}
 \big(V_i^+(h+1/2) - (h+1/2)V_i^-\big)\,.
\end{equation}

Then, making use of the commutators of $h$ with  $V_i^\pm$
(\ref{4.29}) we can easily compute the commutators of $X_i$ and
$P_i$ with a general function $F(h)$ of $h$. Therefore, we obtain
without difficulty that
\begin{eqnarray}
 X_i\,F(h)=
\frac1{2h}\left\{F(h{-}1)(h+1/2)-F(h{+}1)(-h+1/2)\right\}X_i
\\[1.5ex]
\hskip 1cm -
\frac{i}{2h}\,\left\{F(h{-}1)-F(h{+}1)\right\}P_i\,,\nonumber
\end{eqnarray}
and
\begin{eqnarray}
 P_i\,F(h)=
\frac{i}{2h}\,\left\{F(h{-}1)(h^2-1/4)-F(h{+}1)(h^2-1/4)\right\}X_i
\nonumber
\\[1.5ex]
\hskip 1cm +\frac1{2h}\,
\left\{F(h{-}1)(h-1/2)+F(h{+}1)(h+1/2)\right\}P_i\,. \label{4.32}
\end{eqnarray}

  We have already computed the commutators of $h$ with
$V_i^\pm$ and, since the symmetry algebra of $H$ is spanned by
$J_{ij}$, we have $[h,J_{ik}]=0$. Thus, the only commutators that
remain to be found are those involving the eigenoperators $V_i^\pm$.
After some lengthy but straightforward calculations, we find

\begin{equation}\label{4.33}
 [V_i^-,V_j^+]= 2h\delta_{ij} + 2iJ_{ij}\,,\quad
  [V_i^+,V_j^+]=[V_i^-,V_j^-]=0\,.
\end{equation}

At this point, we should remark that the eigenoperators $V_i^\pm$
given in (\ref{4.23}) are defined up to a factor that can depend on
$h$, so that we may replace $V_i^\pm$ by expressions like
 $f(h)V_i^\pm g(h)$. This fact can be used, for
instance, to construct a new pairs of eigenoperators each one
adjoint of the other. Then, since
\begin{equation}\label{4.34}
   (V_i^+)^\dagger=\frac{h+1}{h}\,V_i^-\,,
\end{equation}
we can define
\begin{equation}\label{4.35}
A_i^\pm:=  \frac1{\sqrt{h}}\, V_i^\pm\,\sqrt{h}\,,\qquad i=1,2,3,4\,
,
\end{equation}
so that
\begin{equation}\label{4.36}
   (A_i^+)^\dagger=A_i^-\,.
\end{equation}
Then, we can restate the relevant relations involving $V_i^\pm$ in
terms of $A_i^\pm$. We begin with (\ref{4.29}):

\begin{equation}\label{4.37}
[h,A^\pm_i] = \pm A^\pm\,.
\end{equation}
Next, we can express $X_i$ and $P_i$ as in (\ref{4.30}) in the
following form:
\begin{equation}\label{4.38}
 X_i=\frac{i}{2}\frac1{\sqrt{h}}\;(A_i^+-A_i^-)\frac1{\sqrt{h}}\,,\qquad
P_i=-\frac{1}{2}\frac1{\sqrt{h}}
 \;\big(A_i^+(h+1/2) + (h+1/2)A_i^-\big)\frac1{\sqrt{h}}\,.
\end{equation}
The last formula in (\ref{4.38}) may also be written as

\begin{equation}\label{4.39}
P_i=-\frac{1}{2}\,\frac{f(h)}{\sqrt{2h}}
 \;(A_i^+ + A_i^-)\frac{f(h)}{\sqrt{2h}}\,,
\end{equation}
where the function $\phi$ is characterized by

\begin{equation}\label{4.40}
2h+1 =f(h)f(h+1) \, .
\end{equation}
Equation (\ref{4.40}) is identical to (\ref{3.49}) and therefore it
has the same solution (\ref{3.50}). Finally, commutation relations
(\ref{4.33}) are preserved:

\begin{equation}\label{4.41}
 [A_i^-,A_i^+]= 2h \delta_{ij}+  2iJ_{ij}\,,\quad
  [A_i^+,A_j^+]=[A_i^-,A_j^-]=0\,.
\end{equation}

Now, we are in position to obtain the generators of the
$\frak{so}(4,2)$ Lie algebra from the above ingredients. Then, we
can define the following new operators

\begin{equation}\label{4.42}
 L_i:= \frac1{2i}\,(A_i^--A_i^+)\,,\quad {\rm and}\quad
{K}_i:= \frac12\,(A_i^-+A_i^+)\,.
\end{equation}
Using the commutation relations obtained before in this Section, it
is easy to show that set of the operators

\begin{equation}\label{4.43}
\{ J_{ij}, h, L_i, K_i\}
\end{equation}
form a basis for $\frak{so}(4,2)$. In fact, if we make the following
identifications
\begin{equation}\label{4.44}
 J_{ij}\equiv M_{ij}\,,\quad  {K}_i\equiv M_{5i}\,,\quad
L_i\equiv
  M_{6i}\,,\quad h\equiv M_{56}\,, \quad i,j=1,\dots,4\,,
\end{equation}
then, $M_{\alpha\beta}$, $\alpha,\beta=1,\dots,6$ will satisfy the
commutation relations (\ref{3.1}). The vector character of the
components $L_i$ and $K_i$ or the scalar behaviour of $h$ under
$\frak{so}(4)$ are satisfied from its very definition in terms of
the vectors $X_i$, $P_i$ and the scalar $H$. Notice that, according
to (\ref{4.38}) and (\ref{4.39}), we have

\begin{equation}\label{4.45}
K_i = - \frac{\sqrt{2h}}{f(h)}\, P_i\,\frac{\sqrt{2h}}{f(h)},\qquad
L_i = \sqrt{h}\, X_i \sqrt{h}\, .
\end{equation}
We can avoid the use of the $f(h)$ function in in the relation
between $P_i$ and $K_i$ as follows. For instance, take the second
equation in (\ref{4.38}) and re-express it in the form
\begin{equation}\label{4.46b}
 P_i=-\frac1{\sqrt h}\,\frac14\,[h(A_i^-+A_i^+)+
 (A_i^-+A_i^+)h]\,\frac1{\sqrt h} =
 -\frac12\,\frac1{\sqrt h}\, (h{K}_i+{K}_ih)\,\frac1{\sqrt
 h}\,.
\end{equation}
These relations, together with $h = \sqrt{H-1}$ (\ref{4.24}),
constitute the quantum analogue of the classical ones given by
(\ref{221}).

We have already mentioned in (\ref{4.26}) that the spectrum of the
operator $h$ is the set of natural nubers $n=1,2,3,\dots$ while the
spectrum of the Hamiltonian $H$ is given by $n^2-1$. These energy
levels can be connected by a ladder representation, which follows
from the nature of the operators $A_i^\pm$ and that is identical to
the representation given in section 3.3. This shows that
$\frak{so}(4,2)$ is the SGA for the quantum particle in $S^3$.

\subsection{The restrictive relations.}

In Section 3, we have started directly with the quantization of the
algebra $\frak{so}(4,2)$, including the operators $A_i^\pm$. Then,
some naturally chosen restrictive conditions allowed us to define
the generators $X_i$ of the homogeneous space (and their
corresponding momenta $P_i$) satisfying  commutation relations
(\ref{4.1}). Now, we will finish by showing that the above
construction yields naturally to the restrictive relations as given
by (\ref{3.23}).

In the previous subsection we have shown that the quantum SGA was a
Lie algebra obtained from the quadratic algebra generated by
$\{X_i,P_j,J_{ij},H\}$ after a (generalized) change of basis.
As the starting generators were not independent so there must be
certain relations involving the Lie algebra generators that will
turn into the restrictive conditions. In fact, in order to make all
the computations in the above subsection we have made use of two
relations: (i) the expression of $J_{ij}$ in terms of $X_i,P_j$, as
given in (\ref{4.2}) that is equivalent to the vanishing of the
second Casimir (\ref{4.15}) of $\frak{so}(4)$, and (ii) the choice
of $P_i$ given in terms of $J_{ij}$ and $X_j$ in (\ref{4.11}).

Let us start with the relation (ii)  where we substitute $P_i$ in
terms of $K_i$ by means of (\ref{4.46b}), $X_j$ in terms of $L_j$ as
shown in (\ref{4.45}), to get
\begin{equation}\label{4.47}
J_{ik}\, \frac1{\sqrt h}\,L_k\,\frac1{\sqrt h}+ \frac1{\sqrt
h}\,L_k\,\frac1{\sqrt h}\,J_{ik}=\frac1{\sqrt h}\,(h{
K}_i+{K}_ih)\,\frac1{\sqrt
 h}\,.
\end{equation}
Since $h$ and $J_{ik}$ commute,  equation (\ref{4.47}) becomes:
\begin{equation}\label{4.48}
  J_{ik}L_k+L_kJ_{ik}- h{K}_i-{K}_ih=0\,.
\end{equation}
Now taking into account the identification with the generators
$M_{\alpha\beta}$ of $\frak{so}(4,2)$, this equation is just
$\widetilde T_{6i}=0$, thus obtaining the first restrictive
relation. The remaining relations  are obtained by commuting
(\ref{4.48}) with other operators. For instance, let us commute
(\ref{4.48}) with $h$, then we find a second restrictive relation:
\begin{equation}\label{4.49}
J_{ik}{K}_k+ {K}_k J_{ik}+hL_i+L_ih=0 \qquad {\rm or} \qquad
\widetilde T_{5i}=0\,.
\end{equation}

The next one can be obtained by commuting $L_i$ with (\ref{4.48}),
summing over $i$ and taking into account  that
\begin{equation}\label{4.50}
  [L_i,J_{ik}]=-3iL_k\, ,
\end{equation}
then, we arrive to the relation:
\begin{equation}\label{4.51}
 {\bf K}^2-3{\bf L}^2+2h^2+2=0\,.
\end{equation}
Now, we commute (\ref{4.49}) with $K_i$ and sum over $i$ to find:
\begin{equation}\label{4.52}
{\bf L}^2-3{\bf K}^2+2h^2+2=0\,.
\end{equation}
Relations (\ref{4.51}) and (\ref{4.52}) together yield
\begin{eqnarray}
 {\bf K}^2-h^2-1=0 \qquad {\rm or} \qquad
\widetilde T_{55}=0\,, \label{4.53} \\[2ex]
 {\bf L}^2-h^2-1=0 \qquad {\rm or} \qquad
\widetilde T_{66}=0\,. \label{4.54}
\end{eqnarray}

Next, we commute $L_i$ with (\ref{4.49}) and sum over $i$. We
obtain:
\begin{equation}\label{4.55}
{\bf K}\cdot{\bf L}+{\bf L}\cdot{\bf K}=0 \qquad {\rm or} \qquad
\widetilde T_{56}=0 \,.
\end{equation}
The last of the relations $\widetilde T_{ab}=0$ follows by commuting
$L_j$ with (\ref{4.48}). It gives:

\begin{equation}\label{4.56}
 (L_iL_j+L_jL_i)+(K_iK_j+K_jK_i)-(J_{ik}J_{jk}+J_{jk}J_{ik})-2\delta_{ij}=0\,,
\end{equation}
or $\widetilde T_{ij}=0$. Note that the constant $c$ in
(\ref{3.13})-(\ref{3.18}) appears naturally as $c=2$.

The other restrictive relations in (\ref{3.23}), $R^{ab}=0$, can be
obtained as follows: The former, $R^{56}=0$ is just the expression
(\ref{4.15}) stating that the second Casimir vanishes in the chosen
representation. Then, commute (\ref{4.15}) with $L_j$ and $K_i$ to
obtain $R^{6i}=0$ and $R^{5i}=0$ respectively. Then, using the
explicit relation for $R^{5i}$ given in (\ref{3.20}) and commuting
$R^{5i}=0$ with $L_i$, one finally gets the last relation
$R^{ij}=0$. Explicit forms of $R^{ab}$ are given in
(\ref{3.19})-(\ref{3.22}). With the derivation of the restrictive
relations, we conclude the present section.

\section{Concluding remarks.}

In this paper, we have constructed the Spectrum Generating Algebra
(SGA) for the three dimensional sphere $S^3$ in both classical and
quantum cases. Both situations provide a nontrivial problem,
particularly in the quantum case.

In the classical case, we have obtained specific SGA generators
leading to time dependent constants of motion fixing the motion. In
the quantum case, we can study the SGA for $S^3$ from two points of
view. In the former, we start with a quantized version of the
algebra, postulate natural restrictive relations that fixes the
representation and then, obtaining a ladder representation that
connect the whole set of eigenvectors of the free Hamiltonian for
$S^3$, supporting the Hilbert space of an IUR of the SGA algebra.
Finally, we define the position and momentum operators for the
homogeneous ambient space and fix their constraints and gauge
conditions over $S^3$ and their commutation relations.

The second point of view is equivalent and makes the inverse path.
We start with  quantization of the position and momentum operators
by transforming their Dirac brackets into commutators following the
usual recipe established by usual canonical quantization. These
operators determine a representation of the algebra ${\frak so}(4)$.
Then, we construct ladder operators and determine that the SGA for
our situation is ${\frak so}(4)$. Finally, the restrictive relations
postulated in the previous method are obtained as a consequence of
our hypothesis.

This research as interest by itself, although we expect this work to
serve as a preparation for the construction of the SGA of non
trivial potentials in $S^3$ as well as in a spaces with negative
curvature $H^3$, under the same optics.

\section*{Acknowledgements}

 Partial financial support is indebt to the Spanish
Junta de Castilla y Le\'on (Project GR224) and Ministry of Science
of Spain (Project MTM2009-10751), and to the Russian Science
Foundation (Grants 10-01-00300 and 09-01-12123).

\end{document}